\documentclass[prl,twocolumn,groupedaddress]{revtex4}
\usepackage{graphicx}
\begin{document}


\title{Will spin-relaxation times in molecular magnets permit quantum information
processing?}


\author{Arzhang Ardavan}
\author{Olivier Rival}
\author{John J.L. Morton}
\author{Stephen J. Blundell}
\affiliation{Clarendon Laboratory, Department of Physics,
University of Oxford, OX1 3PU, United Kingdom}
\author{Alexei M. Tyryshkin}
\affiliation{Department of Electrical Engineering, Princeton University, Princeton, New Jersey 08544, United States}
\author{Grigore A. Timco}
\author{Richard E.P. Winpenny}
\affiliation{Department of Chemistry, University of Manchester, Oxford Road, Manchester, M13 9PL, United Kingdom}


\date{\today}

\begin{abstract}
Using X-band pulsed electron spin resonance, we report the intrinsic spin-lattice ($T_1$) and phase coherence ($T_2$) relaxation times in molecular nanomagnets for the first time. In Cr$_7M$ heterometallic wheels, with $M$ = Ni and Mn, phase coherence relaxation is dominated by the coupling of the electron spin to protons within the molecule. In deuterated samples $T_2$ reaches 3~$\mu$s at low temperatures, which is several orders of magnitude longer than the duration of spin manipulations, satisfying a prerequisite for the deployment of molecular nanomagnets in quantum information applications.
\end{abstract}


\maketitle

Certain computational tasks can be efficiently implemented using quantum logic, in which the information-carrying elements are permitted to exist in quantum superpositions~\cite{nielsen-chuang}. To achieve this in practice, a physical system that is suitable for embodying quantum bits (qubits) must be identified.  Some proposed scenarios employ electron spins in the solid state, for example phosphorous donors in silicon~\cite{kane}, quantum dots~\cite{loss-divincenzo}, heterostructures~\cite{vrijen} and endohedral fullerenes~\cite{harneit,bangbang}, motivated by the long electron-spin relaxation times exhibited by these systems.  An alternative electron-spin based proposal exploits the large number of quantum states and the non-degenerate transitions available in high spin molecular magnets~\cite{LossNat,spiller}. Although these advantages have stimulated vigorous research in molecular magnets~\cite{winpenny-cr-ring-qip-prl,sjbflp,hill-science}, the key question of whether the intrinsic spin relaxation times are long enough has hitherto remained unaddressed.  Here we show, using pulsed electron spin resonance experiments on heterometallic wheels, that the relaxation times in molecular magnets can significantly exceed the duration of coherent manipulations, a prerequisite for the deployment of these systems in quantum information applications.

Molecular magnets comprising clusters of exchanged coupled transition metal
ions have been studied extensively in recent years~\cite{molmag-review}.
They can exhibit a substantial ground state spin with a large and negative
zero-field splitting (ZFS), leading to a spontaneous magnetic moment
parallel to the easy axis. In the absence of a magnetic field, the
configurations in which the moment is `up' or `down' relative to the easy
axis are degenerate, and this bistable nature has stimulated interest in
the application of magnetic clusters as classical~\cite{sessoli} or
quantum~\cite{LossNat,spiller,hill-science,winpenny-cr-ring-qip-prl} information
elements.

Molecules in this class have been synthesised with widely varying
properties, from the $S=10$ highly axial
Mn$_{12}$-acetate~\cite{mn12-synth}, to the diamagnetic ring
Cr$_8$F$_8$Piv$_{16}$~\cite{cr8-synth,cr8-magnetic}. 
A key recent chemical advance
is the the development of procedures for magnetically `doping' a
diamagnetic cluster to synthesise paramagnetic molecules in a systematic
and controllable way~\cite{cr7m-synth}. Thus, substituting a Cr$^{3+}$
($s=3/2$) by a Mn$^{2+}$ ($s=5/2$) or a Ni$^{2+}$ ($s=1$) generates the
$S=1$ Cr$_7$Mn or the $S=1/2$ Cr$_7$Ni respectively. 

Many clusters have been investigated using thermodynamic probes such as
magnetization~\cite{mol-mag-magnetization-review} and heat
capacity~\cite{fominaya-mn12-heatcap}, and spectroscopic probes such as
neutron scattering~\cite{mirebeau-Mn12-neutrons,hennion-Mn12-neutrons,cr7m-ins} and electron spin resonance~\cite{mol-mag-esr-review,mcinnes-spectroscopy}. These strategies have been very successful in determining the energy spectra and magnetic structures, but apart from a limited number of observations of
`demagnetisation tunnelling' there are few~\cite{wernsdorfer-cr7ni} reports of definitive
measurements of relaxation times. However, the feasibility of many of the
most interesting proposed applications, in particular those involving
classical or quantum information processing, is critically dependent on
the intrinsic spin-lattice ($T_1$) and 
phase coherence ($T_2$) times.



Measurements were performed using commercial Bruker Elexsys 580 X-band pulsed ESR spectrometers, employing $^4$He flow cryostats for temperature control. The relaxation times were obtained using standard techniques~\cite{schweiger-jeschke}: $T_2$ from the decay of a 2-pulse Hahn echo sequence,
\begin{equation} 
\pi/2-\tau-\pi-\tau-\mathrm{echo},
\label{twopulse}
\end{equation}
with $\tau$ varying; $T_1$ from the recovery of the magnetisation (measured with a spin echo) after an inversion pulse, 
\begin{equation}
\pi-T-\pi/2-\tau-\pi-\tau-\mathrm{echo},
\label{threepulse}
\end{equation}
with $T$ varying and $\tau$ fixed and short.

Two factors conspire to complicate the measurement of intrinsic lifetimes
in anisotropic magnetic clusters. Firstly, in a crystal the magnetic cores
are typically separated by rather small distances of the order of 1~nm to
10~nm, and are therefore coupled by dipolar interactions. For two free
electron spins, the dipolar interaction is of the order of $100/r^3$
MHz$\cdot$nm$^3$~\cite{schweiger-jeschke}, so this dipolar coupling is the
dominant relaxation mechanism limiting $T_2$ in crystals of clusters.
Secondly, clusters tend to exhibit strongly axial behaviour with
significant zero-field splittings; this lifts the degeneracy of the
$\Delta m_S = \pm1$ transitions in high spin ($S>1/2$) molecules, and
leads to a strong dependence of the transition energies on the orientation
with respect to the external magnetic field. Thus if we are to study a
particular transition in an ensemble of identical molecules, they must be
orientationally ordered. The standard approach to solving the first
problem, dissolving the crystal (thereby increasing the average separation
of the clusters), leads to a second problem, an ensemble of randomly
oriented, highly axial molecules.

We chose the two compounds studied here with these factors in mind.
Cr$_7$Ni has a ground state spin of $S=1/2$, so it exhibits a single ESR
transition and, therefore, no zero-field splitting; the anisotropy of the
$g$-factor, which is small, is the only contribution to a dependence of
the transition energy on external magnetic field orientation. A dilute
dissolved sample of this material is therefore amenable to ESR
measurements without causing problems associated with the orientational
disorder. Cr$_7$Mn is a very closely related compound with a ground state
spin of $S=1$; the similarity in its structure leads us to expect that it
should share relaxation mechanisms with Cr$_7$Ni, but its higher spin
allows us to examine the consequences of the zero-field splitting. For
example, a modulation of the zero-field splitting, which might occur as a
result of coupling with thermally excited mechanical deformations of the
molecule, provides further spin-lattice and phase-coherence relaxation
mechanisms~\cite{zfs-relax}.

Samples were prepared as reported elsewhere~\cite{cr7m-synth}, and
dissolved in toluene. The solution was diluted progressively until the
results reported below were no longer dependent on concentration,
indicating that the dipolar coupling between clusters had become
negligible; this occurred for concentrations below approximately
0.2~mg/ml, corresponding to a mean separation of clusters in excess of
about 25~nm.

\begin{figure}
\includegraphics[width=8cm]{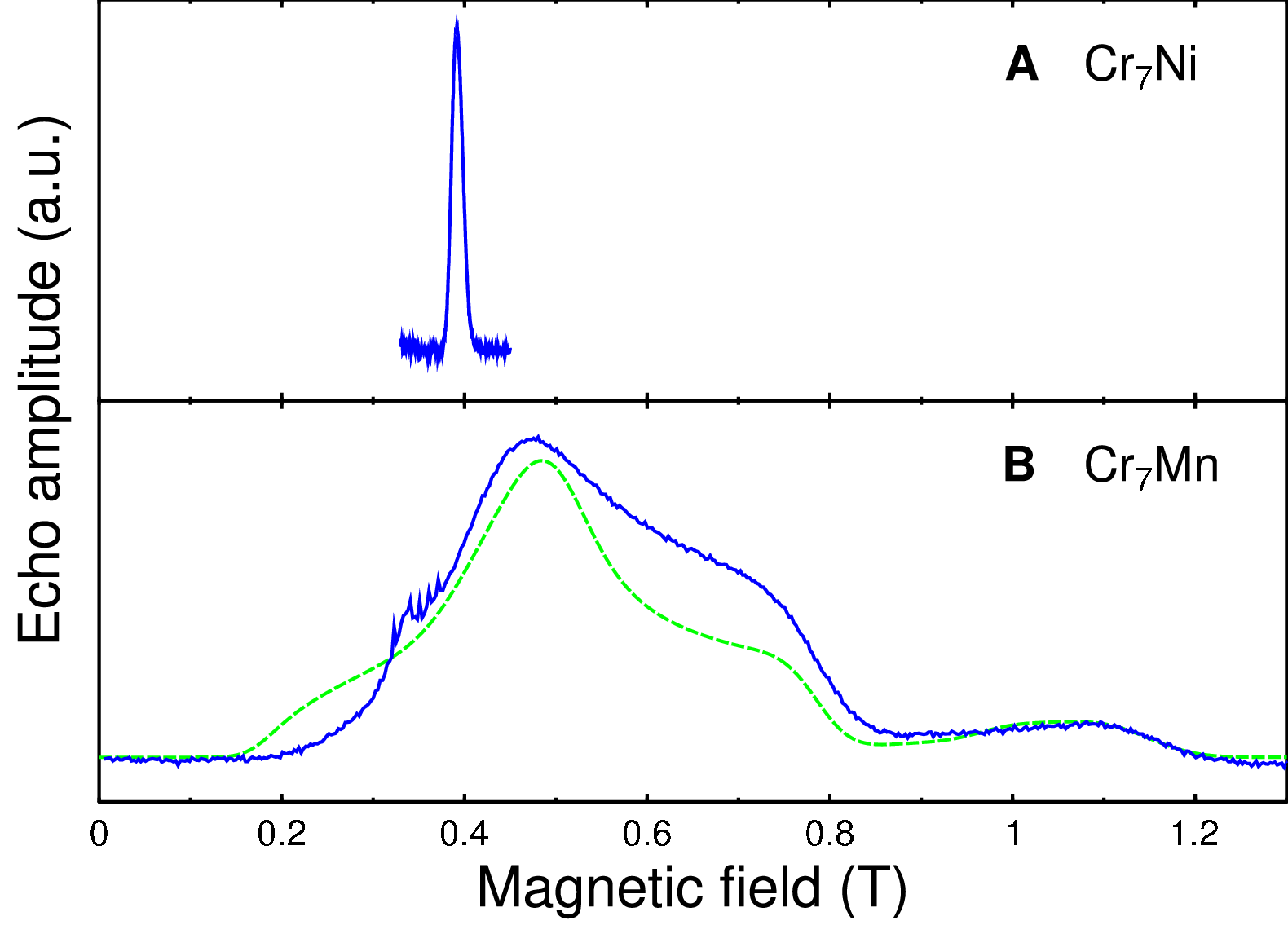}
\caption{(Color online) X-band echo-detected ESR as a function of magnetic field for (A) Cr$_7$Ni and (B) Cr$_7$Mn measured at 4.5~K (blue), and the simulated powder spectrum for a species with $S=1$, $g=2$, $D=21$~GHz, $E=1.9$~GHz (green). In this experiment, the intensity of a Hahn-echo signal with short ($\tau = 300$~ns) delays is measured as a function of the applied magnetic field. Selective pulses, of 64~ns for a $\pi/2$-pulse and 128~ns for a $\pi$-pulse, ensure that only spins within a window of about 0.3~mT are excited. Using a broad integration window suppresses $^1$H ESEEM effects. The echo intensity is proportional to the ESR excitation spectrum. The fine structure in the data close to 0.33~T is an artifact of the cavity.}
\end{figure}

Figure~1
shows echo-detected ESR as a function of the applied magnetic field
measured at 4.5~K for (A) Cr$_7$Ni and (B) Cr$_7$Mn. The spectrum for
Cr$_7$Ni shows a single narrow line, as expected for an $S=1/2$ species.
The line shape is approximately gaussian, with a width of about 0.01~T,
suggesting that the broadening is due to inhomogeneity of the cluster
environment. In contrast, the spectrum for Cr$_7$Mn is broad and contains
structure that is characteristic of an $S=1$ species with a zero-field
splitting exceeding the microwave energy. Also shown is a simulated powder
spectrum~\cite{easyspin} for an $S=1$ species with the Hamiltonian
\begin{equation}
\mathcal{H} = g\mu {\bf B}\cdot{\bf S} + D S_z^2 + E(S_x^2 - S_y^2)
\end{equation}
with parameters $g=1.9$, $D=21$~GHz and $E=1.9$~GHz, which reproduces the
main features of the data (green line in Figure~1(B));
deviations are probably due to the fact that in the experiment, the pulses
deviate from perfect $\pi$ and $\pi/2$ rotations at different parts of the
spectrum because the transition probabilities depend strongly on
orientation. Both spectra show some low-amplitude fine structure close to
0.33~T; these lines arise from impurities in the cavity and are present in
the absence of the sample.

\begin{figure}
\includegraphics[width=8cm]{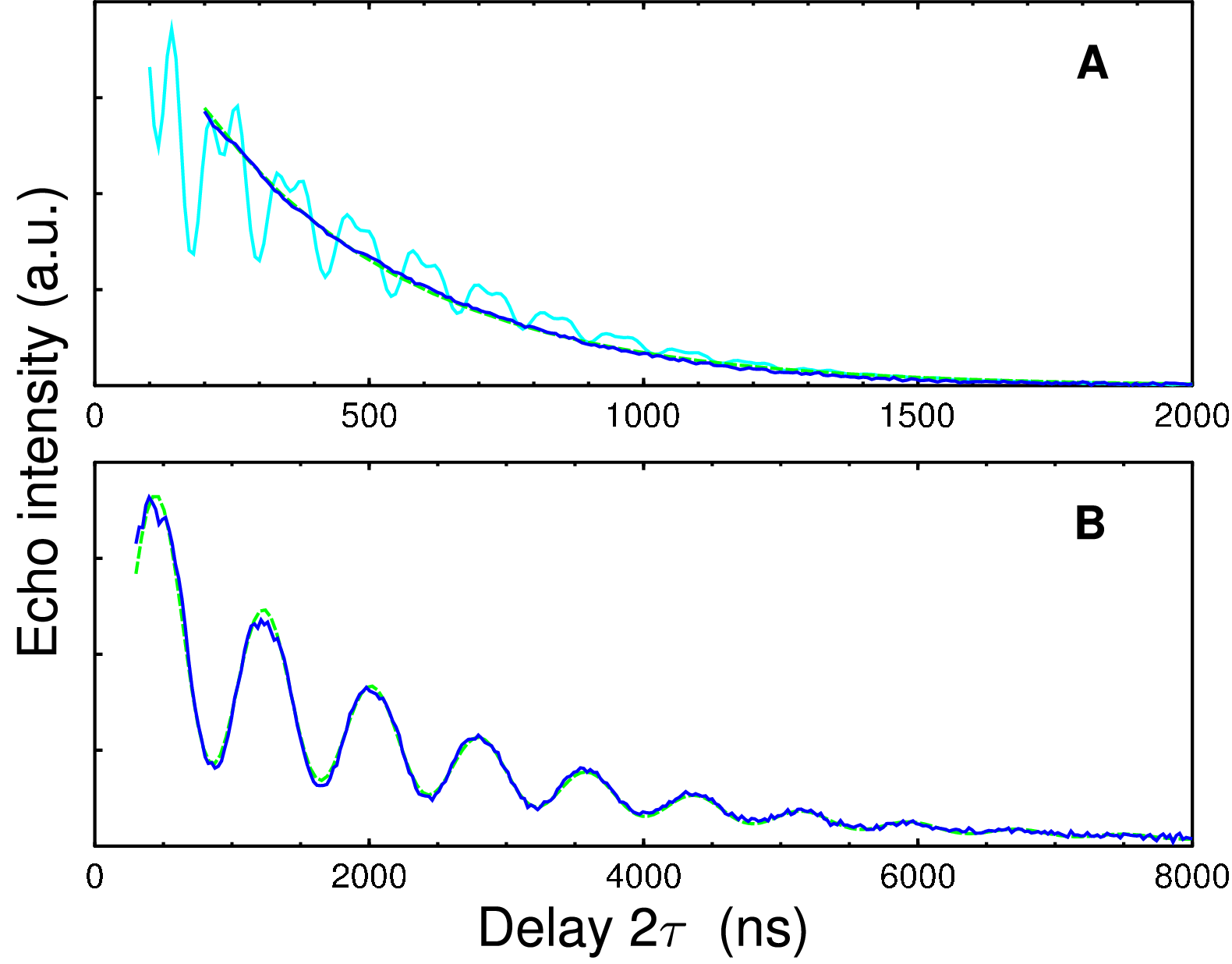}
\caption{(Color online) (A) Decay of the Hahn echo intensity in Cr$_7$Ni as a function of delay time for long, selective pulses (64~ns $\pi/2$-pulse, 128~ns $\pi$-pulse, blue) and for short broadband pulses (16~ns $\pi/2$-pulse, 32~ns $\pi$-pulse, cyan). (B) Decay of the Hahn echo intensity for long, selective pulses in per-deuterated Cr$_7$Ni. Dashed green lines indicate fits to the data. The fit in (B) assumes that the ESEEM effect is dominated by a single harmonic at the $^2$D Zeeman frequency \cite{Dikanov92}. (Note the different horizontal axis scales.)}
\end{figure}

Figure~2(A) shows the intensity of Hahn echos measured in Cr$_7$Ni as a
function of the delay time $\tau$ (defined in equation~\ref{twopulse}),
for long, selective pulses (blue) and short, broadband pulses (cyan). The
echo generated by broadband pulses exhibits very strong modulations
(electron spin echo envelope modulation, or ESEEM) with some second
harmonic content, associated with coupling of the electron spin to
nearby nuclear moments~\cite{schweiger-jeschke}. The nuclear frequency
extracted from the modulation, $16.6 \pm 0.1$~MHz, indicates that the
relevant nuclei are protons. This ESEEM can be suppressed by using longer,
selective pulses, as shown by the blue line in Figure~2(A); this echo
decay is well-described by a mono-exponential fit (green dashed line),
yielding a coherence time of $T_2 = 379 \pm 1$~ns at 4.5~K.

The strong coupling to protons may provide an efficient phase decoherence
path, reducing $T_2$~\cite{stamp-review}. A direct test of whether this is indeed an important relaxation mechanism is to measure the same Hahn echo decay in the
per-deuterated analogue compound, as shown in Figure~2(B). $^2$D has a
gyromagnetic ratio about six times smaller than $^1$H; the ESEEM frequency
is correspondingly about six times smaller at $2.556\pm0.005$~MHz, and $T_2$
is about six times longer, at $2210\pm20$~ns. This confirms that the
coupling to protons dominates the spin decoherence in the hydrogenated
sample. (Note that we were unable to decrease the bandwidth of the pulses
sufficiently to suppress the lower frequency ESEEM in the deuterated
compound.)

\begin{figure}
\includegraphics[width=8cm]{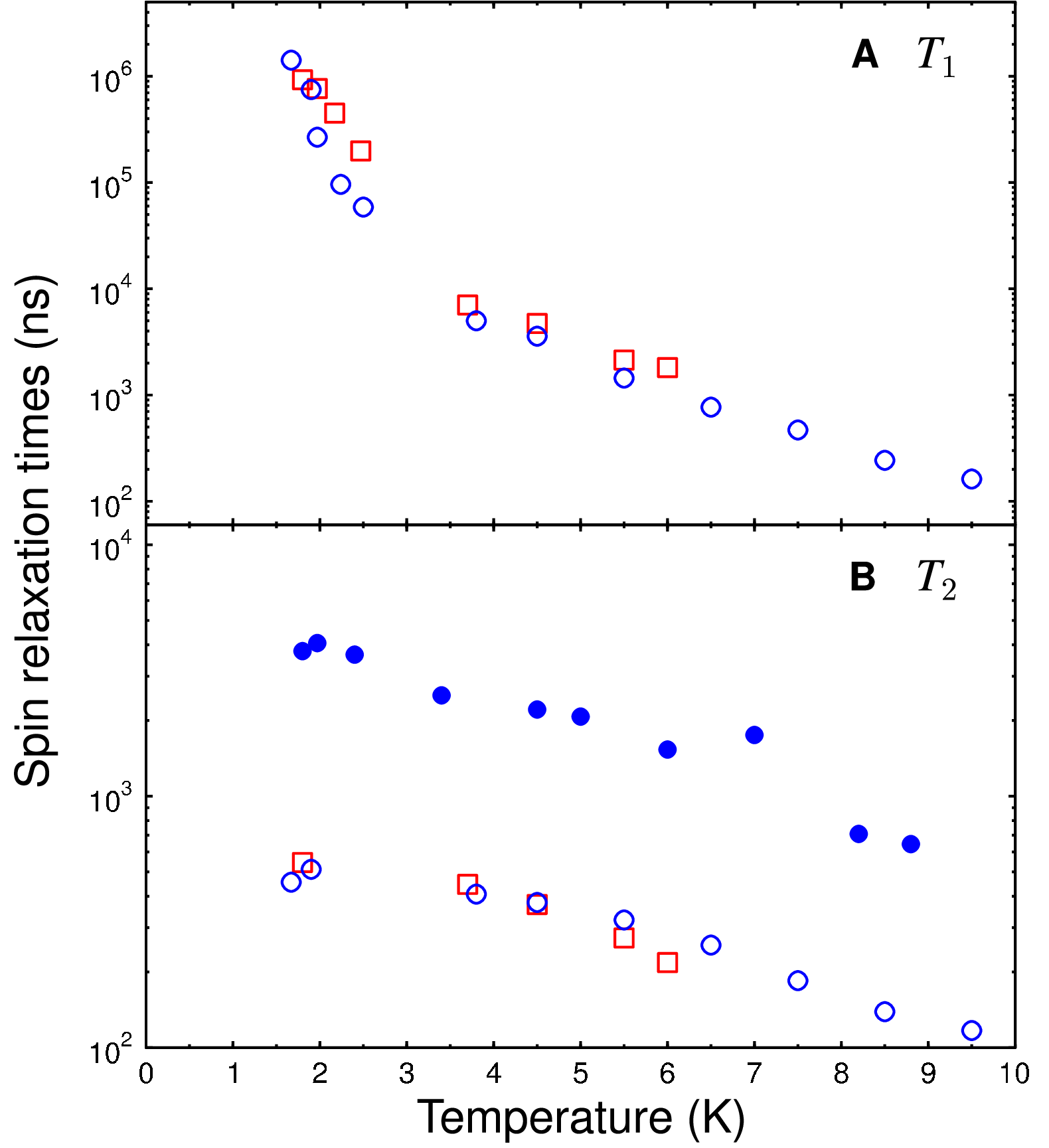}
\caption{(Color online)  (A) $T_1$ as a function of temperature in Cr$_7$Ni (blue open circles) and Cr$_7$Mn (red open squares). (B) $T_2$ as a function of temperature for Cr$_7$Ni (blue open circles), Cr$_7$Mn (red open squares) and per-deuterated Cr$_7$Ni (blue filled circles).}
\end{figure}

Figure~3 shows the temperature dependences of $T_1$ and $T_2$ for each
compound. $T_1$ is comparable between the Cr$_7$Ni and Cr$_7$Mn, and
increases rapidly as the temperature falls. This suggests that thermal
processes (such as couplings to phonons) are responsible for the
longitudinal relaxation. At low temperatures, there is a variation of a
factor of about two in $T_1$ measured at different points in the spectrum
of Cr$_7$Mn. Exciting different parts of the powder spectrum corresponds
to selecting sub-populations of molecules by orientation. These
orientational sub-populations relax at different rates, showing that the
magnetic anisotropy plays at least some r\^ole in the longitudinal
relaxation, probably through spectral diffusion effects~\cite{eatons}.
There is no such variation of $T_2$.

In each compound, $T_2$ also increases as the temperature is decreased,
though less dramatically than $T_1$, and there are signs that it begins to
saturate at temperatures below about 2~K. There are two interesting
observations: first, there is very little difference between the
hydrogenated Cr$_7$Ni and Cr$_7$Mn, despite the strong magnetic anisotropy
of the latter; second, the factor of about six between the decoherence
times for the per-deuterated and hydrogenated Cr$_7$Ni is retained over
the whole temperature range over which a Hahn echo is measurable. Both
observations support the hypothesis that dipolar coupling with $^1$H or
$^2$D nuclei (which belong to organic ligands of the magnetic cluster and
are well-distributed about the core~\cite{cr7m-synth}) dominates the phase
decoherence in these materials. Phase decoherence arising from
fluctuations of the zero field splitting in Cr$_7$Mn is negligible in
comparison over the temperature range studied here.

In conclusion, we have measured intrinsic spin-lattice ($T_1$) relaxation times  and, for the first time (to the best of our knowledge), the phase
coherence ($T_2$) relaxation times in molecular
magnets. We find that in the heterometallic clusters Cr$_7$Ni and
Cr$_7$Mn, $T_1$ is long and somewhat dependent on the magnetic anisotropy
of the cluster, but that $T_2$ is dominated by the coupling to the nuclear
moments of protons in the vicinity of the cluster. There is no evidence of
coupling between the magnetic cluster and the fluorine nuclei, which had
previously been identified as a potential decoherence
path~\cite{winpenny-cr-ring-qip-prl}. Futhermore, we find that the
intrinsic phase coherence time $T_2$ exceeds previous (worst-case)
expectations by three orders of magnitude, reaching 0.55~$\mu$s at
1.8~K, and 3.8~$\mu$s for the per-deuterated analogue. With existing
apparatus, the timescale for coherent manipulations of the electron spin
is of the order of 10~ns; if heteromagnetic clusters of this class were to
be used as elements of a quantum information processing device, this would
lead to a single-qubit figure of merit of several hundred. The
identification of coupling to the $^1$H or $^2$D nuclei as the main
decoherence path offers a strategy for synthesising structures with even
better coherence properties by reducing as far as possible the number of
hydrogens and other magnetic nuclei in the vicinity of the cluster. Our
results are very encouraging for the prospects of constructing and
manipulating non-trivial quantum states within individual
clusters~\cite{LossNat,spiller} and between clusters in
composites~\cite{hill-science}.

We would like to thank C. Kay of University College London, and S.A. Lyon
of Princeton University for the use of their spectrometers and helpful
discussions. This work was supported by EPSRC grants No.\ GR/S57396/01, No.\ GR/T27341/01 and No.\ EP/D048559/1.
A.A.\ is supported by the Royal Society. J.J.L.M.\ is supported by St John's College, Oxford.

\bibliography{Cr7M.bib}

\end{document}